\documentclass[aps,prl,twocolumn,showpacs]{revtex4-1}
\usepackage{epsfig}
\usepackage{bm}
\usepackage{color}

\newcommand{\be}{\begin{eqnarray}}
\newcommand{\ee}{\end{eqnarray}}
\newcommand{\ba}{\begin{array}}
\newcommand{\ea}{\end{array}}
\newcommand{\bmat}{\left(\begin{array}}
\newcommand{\emat}{\end{array}\right)}
\newcommand{\no}{\nonumber}

\begin{document}
\title{Quantum Speed Limit is Not Quantum}
\author{Manaka Okuyama$^1$}
\author{Masayuki Ohzeki$^2$}%
\affiliation{%
$^1$Department of Physics, Tokyo Institute of Technology, Oh-okayama, Meguro-ku, Tokyo 152-8551, Japan
}
\affiliation{%
$^2$Graduate School of Information Sciences, Tohoku University, Sendai 980-8579, Japan
}

\date{\today}

\begin{abstract} 
The quantum speed limit (QSL), or the energy-time uncertainty relation, describes the fundamental maximum rate for quantum time evolution and has been regarded as being unique in quantum mechanics.
In this study, we obtain a classical speed limit corresponding to the QSL using the Hilbert space for the classical Liouville equation.
Thus, classical mechanics has a fundamental speed limit, and
QSL is not a purely quantum phenomenon but a universal dynamical property of the Hilbert space.
Furthermore, we obtain similar speed limits for the imaginary-time Schr\"odinger equations such as the master equation.

\end{abstract}

\maketitle

{\it Introduction.}---
Noncommutativity is one of the most important components of quantum mechanics.
The Heisenberg uncertainty principle \cite{Heisenberg} stems from the canonical commutation relations \cite{Robertson}. Because this consequence cannot appear in classical systems, Heisenberg’s uncertainty principle is a purely quantum phenomenon.
The product of energy and time has the same dimensions as the product of position and momentum, 
which naively implies the existence of a similar relation between energy and time.
However, the time operator, which satisfies the canonical commutation relations for the Hamiltonian, does not exist in a realistic model \cite{Pauli}, and thus, there is no energy-time uncertainty principle that strictly corresponds to Heisenberg's uncertainty principle.
Properly formulating the uncertainty relation for energy and time is a delicate issue that is still being discussed \cite{Hilgevoord,Busch}.

The first rigorous derivation of an analogous uncertainty principle for energy and time was given by Mandelstam and Tamm \cite{MT} in which
they determined that the product of the energy variance and time required for a state to be orthogonal to its initial state was greater than Planck's constant.
This result implied that quantum mechanics has a fundamental speed limit characterized by Planck's constant, and thus, this inequality is called the energy-time uncertainty relation or quantum speed limit (QSL).
The quantum speed limit can be also regarded as a trade-off between energy and time in the variance of a state.
Investigating the restrictions on the time evolution of quantum dynamics is an interesting and important problem, and there are many related works: 
an alternative quantum speed limit \cite{ML}, the shortest time for quantum computation \cite{Lloyd}, cases on mixed states \cite{GLM}, time-dependent systems \cite{Uhlmann,DL,Pfeifer}, and open systems \cite{TEDF,CEPH,DL2}, geometric derivations of the QSL \cite{AA,JK,Zwierz,RS,PCCAP}, and various applications \cite{Bhattacharyya,MP,LT,MMGC,Hegerfeldt,CMCFMGS,Luo,DL3}.	

Note that the QSL is a strictly different concept than Heisenberg's uncertainty principle.
Nevertheless, since QSL appears in a similar context to Heisenberg's uncertainty principle, QSL has been considered a purely quantum phenomenon with no corresponding concept in classical mechanics.
Recent studies \cite{DL, Deffner, GS,GC} have implied that QSL vanishes in the classical limit, and the time evolution of classical mechanics has no fundamental speed limit. However, in this Letter, we show that a fundamental speed limit exists even in classical mechanics.
Inspired by the fact that QSL was obtained from the Hilbert space for the Schr\"odinger equation \cite{MT, ML}, we utilized a similar analysis on the classical Liouville equation \cite{Koopman,JB}.
We rigorously proved that classical mechanics also has a fundamental speed limit, namely, the classical speed limit (CSL).
As a result, we concluded that QSL is not a peculiar phenomenon to quantum mechanics; instead, QSL is a universal property of the time evolution of the Hilbert space.
Furthermore, this method was applied to the imaginary-time Schr\"odinger equation, e.g., the Fokker--Planck equation and the Master equation,
and we showed that these equations also have fundamental speed limits.

{\it  Quantum speed limit.}---
We consider the time-independent Hamiltonian $\hat{H}$. The state $|\phi(t)\rangle$ of the system satisfies the Schr\"odinger equation,
\be
i\hbar\frac{\partial}{\partial t} |\phi(t)\rangle &=&\hat{H} |\phi(t)\rangle.
\ee
Then, the minimal evolution time $\tau_{\text{QSL}}$ needed for the state to rotate orthogonally is bounded as \cite{MT},
\be
\tau \ge \tau_{\text{QSL}} = \frac{\pi \hbar}{2\Delta E}, \label{MT}
\ee
where $\Delta E$ is the energy variance defined as $\sqrt{\langle \phi|\hat{H}^2|\phi\rangle-\langle \phi|\hat{H}|\phi\rangle^2 }$. 
This inequality is known as the Mandelstam--Tamm bound.

Another quantum speed limit, which is called the Margolus--Levitin bound \cite{ML}, is given by 
\be
\tau \ge \tau_{\text{QSL}} = \frac{\pi \hbar}{2  E} , \label{ML}
\ee
where $E$ is the mean energy $\langle \phi|\hat{H}|\phi\rangle$.

In general, these speed limits are independent,  and thus, the minimal evolution time is restricted as follows:
\be
\tau\ge \tau_{\text{QSL}} = \max \left\{ \frac{\pi \hbar}{2\Delta E}  ,\frac{\pi \hbar}{2  E} \right\} .
\ee
See Ref. \cite{GLM} for the case where the two states are not orthogonal.

{\it  Hilbert space for the classical Liouville equation.}---
Consider the time-independent classical  Hamiltonian $H({\bf z})$, where ${\bf z}=(x_1, \cdots, x_N, p_1, \cdots, p_N)$ specifies a point in $N$-dimensional phase space and $H({\bf z})$ belongs to a integrable or nonintegrable system. The evolution of the phase space distribution $\rho({\bf z},t)$ obeys the classical Liouville equation,
\be 
i\frac{\partial \rho({\bf z},t)}{\partial t}  = \hat{L} \rho({\bf x},t) \equiv   i \left\{ H({\bf z}) ,  \rho({\bf z},t)  \right\} ,
\ee
where $\{ \cdot , \cdot \}$ denotes the Poisson bracket, $\hat{L}$ is called the Liouvillian, and $\rho({\bf z},t)$  is normalized as $\int d {\bf z} \rho({\bf z},t)=1$.
The  Liouvillian is known as the Hermitian operator for inner product $\langle \rho_1|\rho_2\rangle\equiv \int d {\bf z} \rho_1^\ast({\bf z}) \rho_2 ({\bf z}) $.
Then, using  the eigenstate $|n\rangle$ of $\hat{L}$, we can  expand  $\rho({\bf z},t)$ as
\be 
|\rho(t)\rangle=\sum_n c_n e^{-i\lambda_n t}|n\rangle,
\ee 
where $c_n$ is a time-independent constant and  $ \langle\rho(t) |\rho(t)\rangle \neq 1$.
We note that the eigenvalues $\lambda_n$ of $\hat{L}$ are symmetric with respect to the origin and $\langle \rho^{}|\hat{L}|\rho^{}\rangle = 0$ (see Supplementary material for the proof).

In the following, we obtain the CSL for the classical Liouville equation by using the Hilbert space for the classical Liouville equation. 

{\it  Margolus--Levitin-type bound for classical Liouville equation.}--
First, we obtain the CSL corresponding to the  Margolus--Levitin bound (\ref{ML}).
We consider the overlap between initial state $|\rho\rangle$ and final state $|\rho(t)\rangle$.
We evaluate $\langle \rho | \rho(t)\rangle$ as
\be
\langle \rho | \rho(t)\rangle&=&\sum |c_n|^2 \cos(\lambda_n t)
\no\\
 &\ge& \sum |c_n|^2 \left(1- \frac{\lambda_n^2 t^2}{2} \right) \label{C-ML-I} ,
\ee
where we use $\cos x\ge1-x^2/2$ and $\langle \rho|\rho(t)\rangle$ takes a real value.
From Eq. (\ref{C-ML-I}), we obtain CSL for the classical Liouville equation,
\be
\tau \ge \tau_{\text{CSL}} = \sqrt{\frac{2 \left(\langle \rho | \rho\rangle- \langle \rho | \rho(t)\rangle \right) }{ \langle \rho| \hat{L}^2|\rho\rangle} }.
\ee
This means that classical time development is also restricted, and there is a trade-off between the classical Hamiltonian (or the Liouvillian) and time during time evolution of the phase space distribution.

From the derivation, this CSL corresponds to the Margolus--Levitin bound in quantum systems \cite{ML,GLM}.
However, we note that we cannot use an odd function for evaluating the inequality because eigenvalues $\lambda_n$ always take symmetric positive and negative values, unlike quantum systems.
For this reason, the form of the CLS is different than the Margolus--Levitin bound (\ref{ML}) in quantum systems.

We stress that QSL and CSL are derived not from noncommutativity; they are dynamical properties of the Hilbert space.
This implies that there are general fundamental speed limits for time-evolution systems. Later, we will show that several stochastic equations have also similar speed limits.

Next, let us determine the single particle limit of CSL. 
For example, we consider the one-dimensional harmonic oscillator,
\be
\rho(x,p,0)&=&\frac{\sqrt{ab}}{\pi}e^{-a(x-e)^2-b(p-f)^2},
\\
H(x,p)&=&d p^2 + c x^2,
\ee
where $\int dx \int dp \rho(x,p)=1$, $d=1/(2m)$, and $c=m\omega^2/2$.
After straightforward calculation, we obtain
\be
\langle \rho|\hat{L}^2|\rho\rangle&=&\frac{(ad-bc)^2+ 4a^2bd^2 f^2+4ab^2 c^2e^2 }{2\pi\sqrt{ab}}.
\ee
Therefore, taking a single particle limit $a,b\rightarrow \infty$, we find that CSL vanishes.
This means that CSL is essentially an effect of many particles in classical systems, which is a natural consequence, because we cannot define the overlap of the phase space distribution for single particles in classical mechanics.

Although $\langle \rho|\rho\rangle$ has not been normalized, we can adjust it  using the fact that the square root of $\rho ({\bf z},t)$ also satisfies the Liouville equation,
\be
i\frac{\partial \rho^{1/2} ({\bf z},t)}{\partial t}  = \hat{L} \rho^{1/2} ({\bf z},t).
\ee
This enables us to expand  $\rho^{1/2}({\bf z},t)$ as $|\rho^{(1/2)}(t)\rangle=\sum_n c_n^{(1/2)} e^{-i\lambda_n t}|n\rangle$, where $c_n^{(1/2)}$ is a time-independent constant and  $\langle \rho^{(1/2)}(t)|\rho^{(1/2)}(t)\rangle = 1$.
Using the same technique as before, we obtain another speed limit for the classical Liouville equation,
\be
\tau \ge \tau_{\text{CSL}}^{(1/2)} =\sqrt{\frac{2 \left(1- \langle \rho^{1/2} | \rho^{1/2}(t)\rangle \right) }{ \langle \rho^{1/2}| \hat{L}^2|\rho^{1/2}\rangle} } .
\ee
Furthermore, we immediately find that $\rho^{\alpha}({\bf z},t)$ satisfies 
\be
i\frac{\partial \rho^{\alpha} ({\bf z},t)}{\partial t}  = \hat{L} \rho^{\alpha} ({\bf z},t),
\ee
and can be expanded as $|\rho^{(\alpha)}(t)\rangle=\sum_n c_n^{(\alpha)} e^{-i\lambda_n t}|n\rangle$, where $\alpha$ is any real value and $c_n^{(\alpha)}$ is a time-independent constant.
Thus, we obtain an infinite number of classical speed limits,
\be
\tau \ge \tau_{\text{CSL}}^{(\alpha)} =  \sqrt{\frac{2 \left(\langle \rho^{(\alpha)}| \rho^{(\alpha)}\rangle- \langle \rho^{(\alpha)} | \rho^{(\alpha)}(t)\rangle \right) }{ \langle \rho^{(\alpha)}| \hat{L}^2|\rho^{(\alpha)}\rangle  } }. \label{CML}
\ee
For a given phase space distribution and Hamiltonian, these speed limits always hold.
Note that we cannot generally compare these inequalities for different parameters between $\alpha$ and $\alpha'$.

{\it  Mandelstam--Tamm-type bound for classical Liouville equation.}--
Next, we obtain the Mandelstam--Tamm-type bound for the classical Liouville equation.
We take the derivative of $\langle \rho^{(\alpha)} | \rho^{(\alpha)}(t)\rangle$ with respect to t,
\be
\left|\frac{d \langle \rho^{(\alpha)} | \rho^{(\alpha)}(t)\rangle}{dt} \right|&=&\left| \sum_n |c_n^{(\alpha)}|^2  \lambda_n \sin(\lambda_n t) \right|
\no\\
&\le& \left|\sum_n |c_n^{(\alpha)}|^2  \lambda_ne^{-i\lambda_n t} \right|
\no\\
&=& \left|\sum_n |c_n^{(\alpha)}|^2  \lambda_n \right.
\no\\
&&\left.  \left(e^{-i\lambda_n t}- \frac{\langle \rho^{(\alpha)} | \rho^{(\alpha)}(t)\rangle} { \langle \rho^{\alpha}|\rho^{\alpha}\rangle } \right) \right|,
\ee
where we use $\sum_n |c_n^{(\alpha)}|^2  \lambda_n =\langle \rho^{(\alpha)} | \rho^{(\alpha)}\rangle=0$ in the last identity.
Furthermore, the Cauchy--Schwarz inequality leads to
\be
\left|\frac{d \langle \rho^{(\alpha)} | \rho^{(\alpha)}(t)\rangle}{dt}  \right|&\le&\sqrt{\langle \rho^{(\alpha)} |\hat{L}^2|\rho^{(\alpha)}\rangle}  
\no\\
&& \sqrt{ \langle \rho^{\alpha}|\rho^{\alpha}\rangle -\frac{\langle \rho^{(\alpha)} | \rho^{(\alpha)}(t)\rangle^2}{ \langle \rho^{\alpha}|\rho^{\alpha}\rangle }   }.	
\ee
Therefore, we obtain another CSL
\be
\tau \ge \tau_{\text{CSL}}^{(\alpha)} =   \frac{\arccos \frac{\langle \rho^{(\alpha)}|\rho^{(\alpha)}(t)\rangle}{ \langle \rho^{(\alpha)}|\rho^{(\alpha)}\rangle } }{ \sqrt{\frac{\langle \rho^{(\alpha)} |\hat{L}^2|\rho^{(\alpha)}\rangle}{\langle \rho^{(\alpha)}|\rho^{(\alpha)}\rangle } } } . \label{CMT}
\ee
Although this CSL corresponds to the Mandelstam--Tamm bound in quantum systems\cite{ML,GLM} from the derivation,
we note that this speed limit cannot represented by  $\Delta\hat{L}$ because $\langle \rho^{(\alpha)}|\hat{L}|\rho^{(\alpha)}\rangle=0$.
This is different from the Mandelstam--Tamm bound in quantum systems.

Although it is not possible to compare the speed limits Eqs. (\ref{MT}) and (\ref{ML}) in quantum systems, we can compare two speed limits Eqs. (\ref{CML}) and (\ref{CMT}) in classical systems. We immediately find 
\be
 \tau \ge  \frac{\arccos \frac{\langle \rho^{(\alpha)}|\rho^{(\alpha)}(t)\rangle}{ \langle \rho^{(\alpha)}|\rho^{(\alpha)}\rangle } }{ \sqrt{\frac{\langle \rho^{(\alpha)} |\hat{L}^2|\rho^{(\alpha)}\rangle}{\langle \rho^{(\alpha)}|\rho^{(\alpha)}\rangle } } } \ge  \sqrt{\frac{2 \left(1- \frac{\langle \rho^{(\alpha)} | \rho^{(\alpha)}(t)\rangle}{\langle \rho^{(\alpha)}|\rho^{(\alpha)}\rangle} \right) }{ \frac{\langle \rho^{(\alpha)}| \hat{L}^2|\rho^{(\alpha)}\rangle }{\langle \rho^{(\alpha)}| \rho^{(\alpha)}\rangle} } } .
 \no\\
\ee
Therefore, in classical system, the Mandelstam--Tamm-type bound (\ref{CMT}) is the better classical speed limit than the Margolus--Levitin-type bound (\ref{CML}).

{\it Speed limit for the imaginary-time Schr\"odinger equation.}---
Although we have considered only the classical Liouville equation, the system has been represented by the Hilbert space.
In other words, if time evolution is expressed using a Hermitian operator, it can be expected that the system would have a similar speed limit.	

First, let us consider the Fokker--Planck equation,
\be
\frac{\partial}{\partial t}P(x,t)=\frac{\partial}{\partial x} \left[\left(2\frac{\partial W(x)}{\partial x} +\frac{\partial}{\partial x} \right)P(x,t)\right] .
\ee
This equation has a stationary solution $P_0(x)=\exp(-2W(x))$.
Using $P(x,t)=\exp(-W(x))\psi(x,t)$, we obtain the imaginary-time Schr\"odinger equation, 
\be
-\frac{\partial}{\partial t} \psi(x,t)&=&\left[-\frac{\partial^2}{\partial x^2} +\left( \frac{\partial W(x)}{\partial x} \right)^2-\frac{\partial^2 W(x)}{\partial x^2}\right]\psi(x,t)
\no\\
&\equiv&\hat{H}_{F} \psi(x,t).
\ee
Then, the ground state of $\hat{H}_{F}$ is given by $\phi_0(x,t)=\exp(-W(x))$, which has the ground state eigenvalue $E_0=0$, and the eigenvalues of the excited states are always positive $E_n>0$.
Using the eigenstates $|\psi_n\rangle$ of $\hat{H}_{F}$, we expand the state $\psi(x,t)$ as $|\psi(t)\rangle=\sum_n d_n e^{-E_n t } |\psi_n \rangle$, where $d_n$ is a time-independent constant. 
Then, we evaluate $\langle \psi|\psi(t)\rangle/\langle \psi|\psi\rangle$ as
\be
\frac{\langle \psi|\psi(t)\rangle}{\langle \psi|\psi\rangle}&=&\sum_n  \frac{|d_n|^2}{\langle \psi|\psi\rangle} e^{-E_n t }
\no\\
&\ge&  \exp\left(-\sum_n \frac{|d_n|^2}{\langle \psi|\psi\rangle} E_n t \right)
\no\\
&=&  \exp\left(-t\frac{\langle \psi|\hat{H}_F|\psi\rangle}{\langle \psi|\psi\rangle}  \right).
\ee
where we use Jensen's inequality.
Therefore, we obtain a speed limit corresponding to the Margolus--Levitin bound (\ref{ML}) for the Fokker--Planck equation:
\be
\tau \ge \tau_{\min} = \frac{ \log \langle \psi|\psi\rangle-\log\langle \psi|\psi(t)\rangle}{\frac{\langle \psi|\hat{H}_{F}|\psi\rangle}{\langle \psi|\psi\rangle}}  \label{CSL-FP}.
\ee

Next, we obtain the speed limit corresponding to the Mandelstam--Tamm bound (\ref{MT})  for the Fokker--Planck equation.
We take the derivative of $\langle \psi|\psi(t)\rangle$ with respect to $t$,
\be
-\frac{d}{dt}\langle \psi|\psi(t)\rangle &=&  \sum_n |d_n|^2  E_ne^{-E_n t }
\no\\
&\le&  \sum_n |d_n|^2  E_ne^{-E_n t/2 } .
\ee
Applying the Cauchy--Schwarz inequality, we get
\be
-\frac{d}{dt}\langle \psi|\psi(t)\rangle &\le& \sqrt{ \langle \psi|\hat{H}_F^2|\psi\rangle } \sqrt{ \langle\psi|\psi(t)\rangle} . \label{dnon0}
\ee
Therefore, we find the speed limit	
\be
\tau \ge \tau_{\min} = 2\frac{ \sqrt{ \langle\psi|\psi\rangle} -\sqrt{ \langle\psi|\psi(t)\rangle}  }{ \sqrt{ \langle \psi|\hat{H}_F^2|\psi\rangle }  }. \label{CSL-FP-MT}
\ee

We note that Eqs. (\ref{CSL-FP}) and (\ref{CSL-FP-MT}) are generally  independent, and thus, the speed limit for the Fokker--Planck equation is given by 
\be
 \tau_{\min} = \max \left( \ \frac{ \log \frac{ \langle \psi|\psi\rangle} {\langle \psi|\psi(t)\rangle}}{\frac{\langle \psi|\hat{H}_{F}|\psi\rangle}{\langle \psi|\psi\rangle}}   , \frac{2-2\sqrt{ \frac{\langle\psi|\psi(t)\rangle}{\langle \psi|\psi\rangle}}  }{ \sqrt{\frac{ \langle \psi|\hat{H}_F^2|\psi\rangle  }{\langle \psi|\psi\rangle} }} \right).
\ee

Finally, we consider the master equation and assume that the detailed balance condition holds,
\be
\frac{d}{dt} P(t)= -\hat{W} P(t) .
\ee
Then, the transition matrix $\hat{W}$ can be represented by a symmetric matrix, and its eigenvalues $\lambda_n$ satisfy  $0=\lambda_0<\lambda_1<\lambda_2<\cdots<\lambda_N$. 
Therefore, using the eigenstates $|n\rangle$ of $W$, we can expand the probability $P(t)$ as $|P(t)\rangle=	\sum_n e_n e^{-\lambda_n t} |n\rangle$ and  obtain the speed limit for the Master equation:
\be
 \tau_{\min} = \max \left(  \frac{ \log \frac{ \langle P|P\rangle} {\langle P|P(t)\rangle}}{\frac{\langle P|\hat{W} |P\rangle}{\langle P|P\rangle}}   , \frac{2-2\sqrt{ \frac{\langle P|P(t)\rangle}{\langle P|P\rangle}}  }{ \sqrt{\frac{ \langle P|\hat{W}^2 |P\rangle  }{\langle P|P\rangle} }} \right).
\ee
This is the fundamental speed limit for the Master equation.
We note that, if the detailed balance condition is broken, we cannot obtain a similar fundamental speed limit because the probability cannot be expanded using the eigenstates of $W$.

{\it Conclusions.}--
We have provided fundamental speed limits for classical systems.
Objects are forbidden from exceeding the speed of light; however,
CSL obtained in this Letter is unrelated to the theory of relativity.
We have established a trade-off between energy and time in the evolution of classical many-body systems.

In the classical Liouville equation, the Margolus--Levitin-type bound and the Mandelstam--Tamm-type bound are not independent, and the Mandelstam--Tamm-type bound is always tighter. This is a remarkable difference from the QSL.
In addition, since the Liouvillian only contains the first derivative, we can obtain an infinite number of CSLs.

We emphasize that QSL is not based on noncommutativity. Instead, the QSL is a universal property of dynamical systems described by a Hermitian operator, which enables similar speed limits to be obtained for the imaginary-time Schr\"odinger equation such as the Master equation.

It is an interesting problem to investigate whether  systems described by the non-Hermitian operator (e.g., the Master equation not satisfying the detailed balance condition) have fundamental speed limits.
Recent studies \cite{TCV,FW,SH,IO} show that a break in the detailed balance condition in the Master equation accelerates relaxation to the steady state, which suggests that the fundamental speed limit may be essentially changed.

The authors thank Keisuke Fujii, Yuri Ishiguro, Kazuya Kaneko, Hidetoshi Nishimori, Tomohiro Sasamoto, and Kabuki Takada for useful discussions.

M. Okuyama was supported by JSPS KAKENHI Grant No. 17J10198.
M. Ohzeki was supported by ImPACT Program of Council for Science, Technology and Innovation (Cabinet Office, Government of Japan) and JSPS KAKENHI No. 16K13849, No. 16H04382.


\onecolumngrid

\def\theequation{S\arabic{equation}}
\makeatletter
\@addtoreset{equation}{section}
\makeatother

\setcounter{equation}{0}

\newpage
\section{Supplemental material for \\
``Quantum Speed Limit is not Quantum''}

\twocolumngrid
\section{Hilbert space for Classical Liouville equation}
In this section, we provide some details about the Hilbert space for the classical Liouville equation.
\subsection{Liouvillian as Hermitian operator}
We show that the Liouvillian is a Hermitian operator.
For simplicity, we consider a one-dimensional system.
We obtain 
\be
&&\langle \rho_1(t)| \left(\hat{L} | \rho_2(t')\rangle \right) 
\no\\
&\equiv&\int\int dxdp \rho_1^\ast(x,p,t) \left( \hat{L}\rho_2(x,p,t) \right) 
\no\\
&=&i \int\int dxdp \rho_1^\ast(x,p,t) \left\{H,  \rho_2(x,p,t)  \right\}
\no\\
&=&i\int\int dxdp  \left(\rho_1^\ast(x,p,t)  \frac{\partial H(x,p)}{\partial x} \frac{\partial \rho_2(x,p,t)}{\partial p}  \right.
\no\\
&&-\left. \rho_1^\ast(x,p,t) \frac{\partial H(x,p)}{\partial p} \frac{\partial \rho_2(x,p,t)}{\partial x}\right)
\no\\
&=&i\int\int dxdp  \left(-\frac{\partial \rho_1^\ast(x,p,t)}{\partial p}  \frac{\partial H(x,p)}{\partial x}  \rho_2(x,p,t) \right.
\no\\
&&\left. +\frac{\partial \rho_1^\ast(x,p,t)}{\partial x}  \frac{\partial H(x,p)}{\partial p}  \rho_2(x,p,t)\right)
\no\\
&=&-i \int\int dxdp \left\{H,  \rho_1^\ast(x,p,t)  \right\} \rho_2(x,p,t) 
\no\\
&=&\int\int dxdp  \left(\hat{L}\rho_1(x,p,t)\right)^\ast \rho_2(x,p,t) 
\no\\
&=& \left(\langle \rho_1(t)|\hat{L} \right) | \rho_2(t')\rangle,
\ee
where we used the fact that the surface term vanishes.
Therefore, the Liouvillian $\hat{L}$ is a Hermitian operator.

\subsection{Proof of $\langle \rho(t)|L | \rho(t)\rangle=0$}
We evaluate $\langle \rho(t)|L | \rho(t)\rangle$ as
\be
&&\langle \rho(t)|L | \rho(t)\rangle
\no\\
&=&i\int\int dxdp  \left(\rho^\ast(x,p,t)  \frac{\partial H(x,p)}{\partial x} \frac{\partial \rho(x,p,t)}{\partial p} \right.
\no\\
&&\left. -\rho^\ast(x,p,t) \frac{\partial H(x,p)}{\partial p} \frac{\partial \rho(x,p,t)}{\partial x}\right)
\no\\
&=&i\int\int dxdp  \left(-\frac{\partial \rho^\ast(x,p,t)}{\partial x}  H(x,p) \frac{\partial \rho(x,p,t)}{\partial p}\right.
\no\\
&&\left. -\rho^\ast(x,p,t)  H(x,p) \frac{\partial \rho(x,p,t)}{\partial x\partial p} \right.
\no\\
&&\left.  +\frac{\partial \rho^\ast(x,p,t)}{\partial p}  H(x,p) \frac{\partial \rho(x,p,t)}{\partial x} \right. 
\no\\
&&\left. +\rho^\ast(x,p,t)  H(x,p) \frac{\partial \rho(x,p,t)}{\partial p \partial x} \right)
\no\\
&=&i\int\int dxdp H(x,p)\left\{ \rho(x,p,t), \rho^\ast(x,p,t) \right\},
\ee
where we used the fact that the surface term vanishes.
Then, if $\rho(x,p,t)$ is a real-valued function, we obtain
\be
\langle \rho(t)|L | \rho(t)\rangle=\sum_n |c_n|^2 \lambda_n=0.
\ee

\subsection{Eigenvalues of the Liouvillian}
The eigenvalue equation of the Liouvillian is given by
\be
\hat{L} \rho_n(x,p)&=&i\frac{\partial H(x,p)}{\partial x}\frac{\partial \rho_n(x,p) }{\partial p}-i\frac{\partial H(x,p)}{\partial p}\frac{\partial \rho_n(x,p) }{\partial x}
\no\\
&=&\left(-\frac{\partial H(x,p)}{\partial x}\frac{\partial g_n(x,p) }{\partial p}+\frac{\partial H(x,p)}{\partial p}\frac{\partial g_n(x,p) }{\partial x} \right) 
\no\\
&&+i\left(\frac{\partial H(x,p)}{\partial x}\frac{\partial f_n(x,p) }{\partial p}-\frac{\partial H(x,p)}{\partial p}\frac{\partial f_n(x,p) }{\partial x} \right)
\no\\
&=&\lambda_n  \left(f_n(x,p) +i g_n(x,p)\right),
\ee
where 
\be
\rho_n(x,p)=f_n(x,p) +i g_n(x,p)
\ee
 is the eigenfunction.
From the above equations, we immediately find that the complex conjugate of $\rho_n(x,p)$ satisfies
\be
\hat{L} \rho_n^\ast(x,p)= -\lambda_n \rho_n^\ast(x,p).
\ee
Therefore, if $\lambda_n$ is an eigenvalue of the Liouvillian, $-\lambda_n$ is also an eigenvalue of the Liouvillian.

\end{document}